\documentclass[prd,twocolumn,superscriptaddress,showpacs,nofootinbib,preprintnumbers]{revtex4}

\usepackage{amsmath}
\usepackage{amsfonts}
\usepackage{graphicx}
\usepackage{dcolumn}
\usepackage{hyperref}
\newcommand{\be}{\begin{equation}}
\newcommand{\ee}{\end{equation}}
\newcommand{\bea}{\begin{eqnarray}}
\newcommand{\eea}{\end{eqnarray}}
\newcommand{\beaa}{\begin{eqnarray*}}
\newcommand{\eeaa}{\end{eqnarray*}}

\newcommand{\nn}{\nonumber \\}
\newcommand{\e}{{\rm e}}

\begin{document}

\tolerance=5000

\title{Unified phantom cosmology:  inflation, dark energy and dark matter under the same standard}

\author{S. Capozziello}
\thanks{Electronic address: capozziello@na.infn.it}
\affiliation{Dipartimento di
Scienze Fisiche, Universit\`{a} di Napoli ``Federico II'' and
INFN, Sez. di Napoli, Compl. Univ. Monte S. Angelo, Edificio N,
Via Cinthia, I-80126, Napoli, Italy}
\author{S. Nojiri}
\thanks{Electronic address: snojiri@yukawa.kyoto-u.ac.jp, \\ nojiri@cc.nda.ac.jp},
\affiliation{Department of Applied Physics, National Defence Academy,
Hashirimizu Yokosuka 239-8686, Japan}
\author{S.D. Odintsov}
\thanks{Electronic address: odintsov@ieec.uab.es}
\affiliation{Instituci\`{o} Catalana de Recerca i Estudis Avan\c{c}ats
(ICREA) and Institut d' Estudis Espacials de Catalunya (IEEC/ICE),
Edifici Nexus, Gran Capit\`{a} 2-4, 08034 Barcelona, Spain}

\begin{abstract}
Phantom cosmology allows to account for  dynamics and matter
content of the universe tracing back the evolution to the
inflationary epoch, considering the transition to the non-phantom
standard cosmology (radiation/matter dominated eras) and
recovering the today observed dark energy epoch. We develop
the unified phantom cosmology where the same scalar plays the role
of early time (phantom) inflaton and late-time Dark Energy.
  The recent transition from decelerating to accelerating phase
is  described too by the same scalar field.
The (dark) matter may be embedded in this scheme,
giving the natural solution of the coincidence problem.
It is explained how the proposed unified phantom cosmology
can be fitted against the observations which opens the way to define
all the important parameters of the model.

\end{abstract}


\maketitle

\noindent
1. According to recent astrophysical data the (constant) effective
equation of state (EOS) parameter $w_{\rm eff}$ of dark energy
lies in the interval: $-1.48<w_{\rm eff}<-0.72$ \cite{HM} (see
very recent comparison of observational data from different
sources in \cite{bagla}, and see also \cite{sahni}).
It is clear that standard $\Lambda$-CDM cosmology is in full agreement with
observations.  Nevertheless, it remains  the possibility that universe
is currently in its
phantom DE phase (for recent study of phantom cosmology, see
\cite{unification,phantom,tsujikawa} and refs therein).
Despite the fact that it remains unclear how decelerating FRW world transformed
to the accelerating DE universe, one can try to unify the early time
(phantom?) inflation with late time  acceleration \cite{unification}.
In fact, the phantom inflation has been proposed
in \cite{inflation}. The unified inflation/acceleration universe
occurs in some versions of modified gravity \cite{modified} as well
as for complicated EOS of the universe \cite{salvatore}
(for recent discussion of similar (phantomic) EOS, see \cite{EOS}
and time-dependent viscosious EOS\cite{viscosity}).

In the present paper we consider unified phantom cosmology with the
account of dark matter. Due to the presence of scalar dependent function
in front of kinetic term, the same scalar field may correspond to the
(phantom) inflaton at very early universe, quintessence at the
intermediate epoch and DE phantom at the late universe.
The recent transition from decelerating phase to the accelerating phase
is naturally described there too.
On the same time it is shown that both phantom phases are stable against
small perturbations, and that coincidence problem may be naturally solved
in our unified model. The equivalent description of the same unified
phenomena via
the (multi-valued) EOS is given too. In the final section we explain how
the proposed unified phantom cosmology can be fitted against the
observations which gives the way to define all the important parameters
of the model.

\noindent
2. Let us start from the following action:
\bea
\label{k1}
S&=&\int d^4 x \sqrt{-g}\Bigl\{\frac{1}{2\kappa^2}R
   - \frac{1}{2}\omega(\phi)\partial_\mu \phi\partial^\mu \phi \nn
&& - V(\phi)\Bigr\} + S_m\ .
\eea
Here $\omega(\phi)$ and
$V(\phi)$ are functions of the scalar field $\phi$ and $S_m$ is
the action for matter field. Without matter, such an action has
been proposed in \cite{uni} for the unification of early-time
inflation and late-time acceleration in frames of phantom
cosmology. We now assume the spatially-flat FRW metric
$ds^2 = - dt^2 + a(t)^2 \sum_{i=1}^3 \left(dx^i\right)^2$.
Let the scalar field $\phi$ only depends on the time coordinate $t$. Then the FRW
equations are given by
\be
\label{any1}
\frac{3}{\kappa^2}H^2 = \rho + \rho_m\ ,
\quad - \frac{2}{\kappa^2}\dot H= p + \rho + p_m + \rho_m\ .
\ee
Here $\rho_m$ and $p_m$ are the energy density and
the pressure of the matter respectively. The energy density $\rho$
and the pressure $p$ for the scalar field $\phi$ are given by
\be
\label{k4}
\rho = \frac{1}{2}\omega(\phi){\dot \phi}^2 + V(\phi)\
,\quad p = \frac{1}{2}\omega(\phi){\dot \phi}^2 - V(\phi)\ .
\ee
Combining (\ref{k4}) and (\ref{any1}), one finds
\bea
\label{any2}
&& \omega(\phi) {\dot \phi}^2 = - \frac{2}{\kappa^2}\dot H -
\left(\rho_m + p_m\right)\ ,\nn &&
V(\phi)=\frac{1}{\kappa^2}\left(3H^2 + \dot H\right) -
\frac{\rho_m - p_m}{2} \ . \eea As usually $\rho_m$ and $p_m$
satisfy the conservation of the energy: \be \label{any3}
\dot\rho_m + 3H\left(\rho_m + p_m\right)=0\ .
\ee
As clear from the first equation  (\ref{any1}),
in case without matter ($\rho_m=p_m=0$), when $\dot H$ is positive,
which corresponds to the phantom phase, $\omega$ should be negative,
that is, the kinetic term of the scalar field has non-canonical sign.
On the other hand, when $\dot H$ is negative, corresponding to the non-phantom
phase,
$\omega$ should be positive and the sign of the kinetic term of the scalar
field is canonical.
If we restrict in one of phantom or non-phantom phase, the function
$\omega(\phi)$ can be absorbed into the field redefinition given by
$\varphi = \int^\phi d\phi \sqrt{\omega(\phi)}$ in non-phantom phase or
$\varphi = \int^\phi d\phi \sqrt{-\omega(\phi)}$ in phantom phase.
%%%%%%%
Usually, at least locally, one can solve $\phi$ as a function of
$\varphi$, $\phi=\phi(\varphi)$.
Then the action (\ref{k1}) can be rewritten as
\be
\label{k1B}
S=\int d^4 x \sqrt{-g}\left\{\frac{1}{2\kappa^2}R \mp \frac{1}{2}\partial_\mu \varphi
\partial^\mu \varphi  - \tilde V(\varphi)\right\} + S_m\ .
\ee
Here $\tilde V(\varphi)\equiv V\left(\phi(\varphi)\right)$.
In the sign $\mp$ of (\ref{k1B}), the minus sign corresponds to the non-phantom phase and the plus one to
the phantom phase. Then both of $\omega(\phi)$ and $V(\phi)$ in the action (\ref{k1})
do not correspond to physical degrees of freedom but
only one combination given by $\tilde V(\varphi)$ has real freedom in each of the phantom or non-phantom phase
and defines the real dynamics of the system.
%%%%%%%
The redefinition, however, has a discontinuity between two phases. When explicitly keeping
$\omega(\phi)$, the two phases are smoothly connected with each other (kind of phase transitions).
%%%%%%%%%%
Hence, the function $\omega(\phi)$ gives only redundant degree of freedom and does not correspond to the
extra degree of freedom of the system ( in the phantom or non-phantom
phase). It plays the important role just in the point
of the transition
between the phantom phase and non-phantom phase.
By using the redundancy of $\omega(\phi)$, in any physcally equivalent model, one may choose,
just for example, $\omega(\phi)$ as
$\omega(\phi)=\omega_0\left(\phi - \phi_0\right)$ with constants $\omega_0$ and $\phi_0$. If we further
choose $\omega_0$ to be positive, the region given by $\phi>\phi_0$ corresponds to the non-phantom phase,
the region $\phi<\phi_0$ to the phantom phase, and the point $\phi=\phi_0$ to the point of the transition
between two phases.
%%%%%%%%%%%%%

First we consider the case that the parameter $w_m$ in the matter
EOS is a constant: $w_m=p_m/\rho_m$. (In principle, such dark
matter may be presented via the introduction of one more scalar
field). Then by using (\ref{any3}), one gets $\rho_m=\rho_{m0}
a^{-3(1+w_m)}$. Here $\rho_{m0}$ is a constant. If $\omega(\phi)$
and $V(\phi)$ are given by a single function $g(\phi)$ as
\bea
\label{any5}
\omega(\phi) &=&- \frac{2}{\kappa^2}g''(\phi) -
\frac{w_m + 1}{2}g_0 \e^{-3(1+w_m)g(\phi)}\ ,\nn V(\phi) &=&
\frac{1}{\kappa^2}\left(3g'(\phi)^2 + g''(\phi)\right) \nn && +
\frac{w_m -1}{2}g_0 \e^{-3(1+w_m)g(\phi)} \ ,
\eea
with a positive
constant $g_0$, we find a solution of (\ref{any1}) or (\ref{any2})
given by
\bea
\label{any6}
&& \phi=t\ ,\quad H=g'(t)\ ,\nn &&
\left(a=a_0 \e^{g(t)}\ ,\quad a_0\equiv
\left(\frac{\rho_{m0}}{g_0}\right)^{\frac{1}{3(1+w_m)}}\right)\ .
\eea
Hence, even in the presence of matter, any required cosmology
defined by $H=g'(t)$ can be realized by (\ref{any5}).

More  generally, one may consider the generalized EOS like
\cite{EOS}: $p_m=-\rho_m + F\left(\rho_m\right)$.
Here $F(\rho)$ is a proper function of $\rho_m$.
Using the conservation of the energy (\ref{any3}) gives
$a=a_0 \e^{-\frac{1}{3}\int \frac{d\rho_m}{F(\rho_m)}}$.
Let us assume the above equation can be solved with respect to $\rho_m$ as
$\rho_m=\rho_m(a)$.
Then if we may choose $\omega(\phi)$ and $V(\phi)$ by a single function
$g(\phi)$ as
\bea
\label{any9}
\omega(\phi)&=&- \frac{2}{\kappa^2}g''(\phi)
    - F\left(\rho_m\left(a_0 \e^{g(\phi)}\right)\right)\ ,\nn
V(\phi)&=&\frac{1}{\kappa^2}\left(3g'(\phi)^2 + g''(\phi)\right)
    - \rho_m\left(a_0 \e^{g(\phi)}\right) \nn
&& + \frac{1}{2} F\left(\rho_m\left(a_0 \e^{g(\phi)}\right)\right)\ ,
\eea
with a positive constant $a_0$, we find a solution of (\ref{any1}) or
(\ref{any2}) again:
\be
\label{any10}
\phi=t\ ,\quad H=g'(t)\quad \left(a=a_0\e^{g(t)}\right)\ .
\ee
Hence, any cosmology defined by $H=g'(t)$ can be realized by (\ref{any9}).

Since the second FRW equation is given by
\be
\label{FRW2k}
p=-\frac{1}{\kappa^2}\left(2\dot H + 3H^2\right)\ ,
\ee
by combining the first FRW equation,  the EOS
parameter $w_{\rm eff}$ looks as
\be
\label{FRW3k}
w_{\rm eff}=\frac{p}{\rho}= -1 - \frac{2\dot H}{3H^2}\ .
\ee

%%%%%%%%%
Now it is common to beleive that about 5 billion years ago, the
deceleration of the universe  has
turned to the acceleration. We now show that the model describing such a transition
could be easily constructed in the present formulation.
%%%%%%%%%
As an example, we consider the model with constant $w_m$  (\ref{any5}).
It is also assumed $w_m>-1$. Choosing $g(\phi)$ as
\be
\label{any11}
g(\phi)=\frac{2}{3\left(w_m + 1\right)}\ln \left(\frac{\phi}{t_s -
\phi}\right)\ ,
\ee
we obtain
\bea
\label{any12}
\omega(\phi)&=&- \frac{4}{3\left(w_m + 1\right)\kappa^2}\left(-
\frac{1}{\phi^2}
+ \frac{1}{\left(t_s - \phi\right)^2}\right) \nn
&&  - \frac{w_m + 1}{2}g_0 \frac{\left(t_s - \phi\right)^2}{\phi^2}\ ,\nn
V(\phi) &=& \frac{1}{\kappa^2}\left\{\frac{4}{3\left(w_m + 1\right)^2}\left(
\frac{1}{\phi} + \frac{1}{t_s - \phi}\right)^2 \right. \nn
&& \left. + \frac{2}{3\left(w_m + 1\right)}\left( - \frac{1}{\phi^2}
+ \frac{1}{\left(t_s - \phi\right)^2}\right)\right\} \nn
&& + \frac{w_m - 1}{2}g_0 \frac{\left(t_s - \phi\right)^2}{\phi^2}\ .
\eea
Then the Hubble rate $H$ is given by
\be
\label{any13}
H=\frac{2}{3\left(w_m + 1\right)}\left(\frac{1}{t} + \frac{1}{t_s - t}\right)
\ .
\ee
Since
\be
\label{any14}
\dot H=\frac{2}{3\left(w_m + 1\right)}\left(-\frac{1}{t^2}
+ \frac{1}{\left(t_s - t\right)^2}\right)\ ,
\ee
the EOS parameter $w_{\rm eff}$  (\ref{FRW3k})
goes to $w_m>-1$ when $t\to 0$ and goes to $-2 - w_m<-1$ at large times. The
crossing $w_{\rm eff}=-1$
occurs when $\dot H=0$, that is, $t=t_s/2$.
Note that
\bea
\label{any15}
\frac{\ddot a}{a}%=\dot H + H^2 \nn
&=&\frac{16t_s}{27 \left(w_m + 1\right)^3\left(t_s - t\right)^2 t^2} \nn
&& \times \left\{ t - \frac{\left(3w_m + 1\right)t_s}{4}\right\}\ .
\eea
Hence, if $w_m>-1/3$, the deceleration of the universe turns to the
acceleration at $t= t_a \equiv \left(3w_m + 1\right)t_s/4$.
The energy density of the scalar field $\phi$ and that of the matter are given
by
\bea
\label{any17}
\rho&=&\frac{4t_s^2}{3\kappa^2\left(w_m + 1\right)^2 \left(t_s - t\right)^2 t^2}
    - g_0\frac{\left(t_s - t\right)^2}{t^2}\ ,\nn
\rho_m &=& g_0\frac{\left(t_s - t\right)^2}{t^2}\ .
\eea
Then if  the coincidence time $t_c$ is defined by
$\rho|_{t=t_c}=\rho_m|_{t=t_c}$, we find
$t_c=t_s - \left\{2t_s^2/3g_0 \kappa^2\left(w_m +
1\right)^2\right\}^{1/4}$.
We may assume $t_s$ could be of the order of the age of the universe,
$t_s\sim 10^{10}$yr$\sim \left(10^{-33}\rm{eV}\right)^{-1}$.
On the other hand $\kappa\sim \left(10^{19}\rm{GeV}\right)^{-1}\sim
\left(10^{28}\rm{eV}\right)^{-1}$.
Then there is a mixing of very large parameter $\kappa$ and small one
$1/t_s$  in (\ref{any12}), which can be unnatural.
%%%%%%%%%%%%%%%%%%%
%%%%%%%%%%%%%%%%%%%

%%%%%%%%
We now show that the above problem of the unnaturalness could be also
avoided in the present formulation.
We now propose
%%%%%%%%
the second example in (\ref{any5}) given by
\be
\label{any19}
g(\phi) = - \alpha \ln \left(1 - \beta \ln \frac{\phi}{\kappa}\right)\ .
\ee
Here $\alpha$ and $\beta$ are dimensionless positive constants.
As we explain soon, we choose $\beta \sim {\cal O}\left(10^{-2}\right)$.
Note that the parameter order of $10^{-2}$  is not  unnatural since, say
$\pi^4\sim {\cal O}\left(10^2\right)$.
Eq.(\ref{any19}) gives the following expressions for $\omega(\phi)$ and
$V(\phi)$:
\bea
\label{any20}
\omega(\phi)&=&-\frac{2\alpha\beta\left(\beta - 1 + \beta\ln
\frac{\phi}{\kappa}\right)}
{\kappa^2\left(1 - \beta\ln \frac{\phi}{\kappa}\right)^2\phi^2} \nn
&&  - \frac{w_m + 1}{2}g_0\left(1 - \beta\ln \frac{\phi}{\kappa}
\right)^{3(w_m + 1)\alpha}\ ,\nn
V(\phi)&=&\frac{\alpha\beta\left(3\alpha\beta + \beta - 1 + \beta\ln
\frac{\phi}{\kappa}\right)}
{\kappa^2\left(1 - \beta\ln \frac{\phi}{\kappa}\right)^2\phi^2} \nn
&& + \frac{w_m - 1}{2}g_0\left(1 - \beta\ln \frac{\phi}{\kappa}
\right)^{3(w_m + 1)\alpha}\ .
\eea
Supposing $g_0\sim {\cal O}\left(\kappa^{-2}\right)$, there does not appear
small parameter like $1/t_s$ in (\ref{any20}).
Now the Hubble rate is given by
$H=\alpha\beta/\left(1 - \beta\ln \left(t/\kappa\right)\right)t$,
which is positive if
\be
\label{any22}
0<t<t_s\equiv \kappa \e^{\frac{1}{\beta}}\ ,
\ee
and has a Big Rip type singularity at $t=t_s$ (which apparently may not
occur due to account of quantum effects \cite{tsujikawa}).
Since $10^{61}\sim \e^{140}$, with the choice $\beta \sim 1/140$, we
obtain
$t_s\sim \kappa\times 10^{61}\sim \left(10^{-33}\,{\rm eV}\right)^{-1}$, whose
order is
that of the age of the present universe.
Then now, due to the property of the exponential function (or logarithmic
function),
the small scale like $t_s$ appears rather naturally.
We should also note that if  $\alpha\beta\sim {\cal O}(10^{0-2})$ and
$t$ is a present
age of the universe $t\sim \left(10^{-33}\,{\rm eV}\right)^{-1}$, the observed
value of the Hubble rate $H\sim 10^{-33}\,{\rm eV}$ could be also reproduced.
Since
\be
\label{any23}
\frac{\ddot a}{a}=\frac{\alpha \beta^2\left(\ln\frac{t}{\kappa} + \alpha + 1 -
\frac{1}{\beta}\right)}
{\left(1 - \beta \ln \frac{t}{\kappa}\right)^2 t^2}\ ,
\ee
the universe turns to the acceleration from the deceleration when
\be
\label{any24}
t=t_a \equiv \kappa\e^{\frac{1}{\beta} - \alpha -1}<t_s\ .
\ee
Since the energy density of the scalar field $\phi$ and that of the matter are
given by
\bea
\label{any25}
\rho&=& \frac{3 \alpha^2 \beta^2}{\kappa^2 \left(1 - \beta \ln
\frac{t}{\kappa}\right)^2 t^2} \nn
&& - g_0\left(1 - \beta\ln \frac{t}{\kappa}\right)^{3(w_m + 1)\alpha}\ , \nn
\rho_m &=& g_0\left(1 - \beta\ln \frac{t}{\kappa}\right)^{3(w_m + 1)\alpha}\ ,
\eea
the coincidence time $t_c$ could be given by solving the following equation:
\be
\label{any26}
\left(1 - \beta\ln \frac{t_c}{\kappa}\right)^{3(w_m + 1)\alpha+ 2}t_c^2
=\frac{3\alpha^2\beta^2}{\kappa^2 g_0}\ .
\ee
One may regard $\rho_m$ as the sum of the energy density of usual matter, like
baryons, and
that of (cold) dark matter. If $\rho$ corresponds to the energy density of the
dark energy,
the current data indicate that $\rho:\rho_m\sim 7:3$. Then in the present
universe,
it follows
\be
\label{any27}
\left(1 - \beta\ln \frac{t}{\kappa}\right)^{3(w_m + 1)\alpha+ 2}t^2
\sim \frac{9\alpha^2\beta^2}{10\kappa^2 g_0}\ .
\ee
Thus, in the model (\ref{any19}), the acceleration of the
present universe and the coincidence problem seem to be explained rather naturally.

%%%%%%%
Besides the present acceleration of the universe, there was a period of the accelerated expansion of
universe, which is the inflation of the early universe.
%%%%%%%
One can  further extend the model (\ref{any19}), as the third example, to
explain on the same time also the inflation of the early universe.
By introducing a new dimensionless positive constant $\gamma$,
    the following $g(\phi)$ can be proposed:
\be
\label{any28}
g(\phi)= - \alpha \ln \left( 1 - \frac{\beta}{2}\ln \left(\gamma
+ \frac{\phi^2}{\kappa^2}\right)\right)\ .
\ee
As in (\ref{any19}), $\alpha$ and $\beta$ are dimensionless positive constants
and it is assumed $\beta\sim {\cal O}\left(10^{-2}\right)$.
In the limit  of large $t$ ($t^2/\kappa^2\gg \gamma$) being still less than
$t_s$ in (\ref{any22}),
$g(t)$ coincides with that in (\ref{any19}).
Since the scale factor $a$  is given by  $a=a_0\e^{g(t)}$ as in (\ref{any6}),
the
universe is invariant under the time reversal $t\to -t$ in (\ref{any28}). Then
the universe is shrinking
when $t<0$ and expanding when $t>0$. The scale factor has minimum at $t=0$ as
$a = \left(1 - \left(\beta/2\right)\ln \gamma \right)^{-\alpha}$.
In the model (\ref{any28}), the Hubble rate is given by
\be
\label{any30}
H(t)=\frac{\alpha\beta t}
{\kappa^2 \left( 1 - \frac{\beta}{2}\ln \left(\gamma +
\frac{t^2}{\kappa^2}\right)\right)
\left(\gamma + \frac{t^2}{\kappa^2}\right)}\ .
\ee
Since
\bea
\label{any31}
\frac{\ddot a}{a}&=&\frac{\alpha\beta}
{\kappa^2 \left( 1 - \frac{\beta}{2}\ln \left(\gamma +
\frac{t^2}{\kappa^2}\right)\right)^2
\left(\gamma + \frac{t^2}{\kappa^2}\right)^2} \nn
&& \times \left\{ \left( 1 - \frac{\beta}{2}\ln \left(\gamma +
\frac{t^2}{\kappa^2}\right)\right)
\left(\gamma - \frac{t^2}{\kappa^2}\right) \right. \nn
&& \left. + \frac{\beta(1+\alpha)t^2}{\kappa^2}\right\}\ ,
\eea
if $t>0$, there are two solutions of $\ddot a=0$ , one corresponds to late time
and another corresponds to early time.
The late time solution of $\ddot a=0$ is obtained by neglecting $\gamma$ and
coincides with  (\ref{any24}):
$t=t_l \sim \kappa\e^{1/\beta - \alpha -1}$.
On the other hand, the early time solution could be found by neglecting
$\beta$, which is ${\cal O}\left(10^{-2}\right)$, to be
$t=t_e\sim \kappa \sqrt{\gamma}$.
Then the universe undergoes accelerated expansion when $0<t<t_e$ and
$t_l<t<t_s$.
Here $t_s$ is Rip time:
$t_s=\kappa \sqrt{-\gamma + \e^{2/\beta}}\sim \kappa
\e^{1/\beta}$.
$t_e$ may be identified with the time when the inflation ended.
One is able to define the number of the e-foldings $N_e$ as
$N_e = \ln \left(a\left(t_e\right)/a\left(0\right)\right)$.
Then we obtain
\be
\label{any36}
N_e= -\alpha \ln \left(\frac{1-\frac{\beta}{2}\ln ( 2\gamma )}
{1-\frac{\beta}{2}\ln ( \gamma )}\right)\ .
\ee
%%%%%%%%%%%%%
It is known that $N_e$ should be equal or larger than $60$.
We should note that $\ln$-function in (\ref{any36}) cannot
 be so large naturally since this requires
$\left(1-(\beta/2)\ln ( 2\gamma )\right)/\left(1-(\beta/2)\ln ( \gamma )\right) \sim \e^{60}
\sim 10^{25}$.
Then the $\ln$-function in (\ref{any36}) should be of order of unity,
which  requires that the parameter $\alpha$ should be $\alpha\sim 10^2$.
For example, since we can rewrite (\ref{any36}) as 
$\gamma=\exp\left(2/\beta - \ln 2/(1 - \e^{-N_e/\alpha})\right)$, we find that, when $\alpha=1/\beta=240$, 
we have $N_e=60$ if we choose $\gamma$ as $\gamma=0.043925\cdots$.
%%%%%%%%%%%%%
In the same way one can define number of e-foldings in other similar models.
For instance, in \cite{uni}, without  matter, the following unified model has been considered:
\be
\label{k17}
f(\phi)\equiv g'(\phi)=h_0^2 \left( \frac{1}{t_0^2
- \phi^2} + \frac{1}{\phi^2 + t_1^2}\right)\ .
\ee
Here $h_0$, $t_0$, and $t_1$ are positive constants. It is assumed $t_0>t_1$.
It has been found \cite{uni} that $H$ has two minima at $t=t_\pm
\equiv \pm \sqrt{\left(t_0^2 - t_1^2\right)/2}$ and at $t=0$, $H$
has a local maximum. Hence, (early and late) accelerating phantom
phase occurs when $t_-<t<0$ and $t>t_+$. The number of  e-foldings
may be defined as
\bea
\label{any35b}
N_e &=& \ln
\frac{a\left(0\right)}{a\left(t_-\right)} =-\frac{h_0^2}{2t_0}\ln
\left(\frac{t_0 - \sqrt{\frac{t_0^2 - t_1^2}{2}}} {t_0 +
\sqrt{\frac{t_0^2 - t_1^2}{2}}}\right) \nn
&& + \frac{h_0^2}{t_1}\left({\rm Arctan}\left(t_1\sqrt{\frac{2}{t_0^2 -
t_1^2}}\right) + \pi\right)\ .
\eea
%%%%
Since $\ln$ and ${\rm Arctan}$-functions should be the order of unity, we find
$h_0^2/t_1$ (Note that we have assumed $t_0>t_1>0$.) should be ${\cal O}\left(10^2\right)$ so that
$N_e$ could be equal or larger than $60$. 
For example, when $t_0\gg t_1$, we find $N_e\sim \frac{h_0^2\pi}{t_1}$. Then if we choose 
$\frac{h_0^2\pi}{t_1}\sim 60$, we find $N_e\sim 60$. 
%%%%
Hence, it is demonstrated that scalar field may play the role of phantom
inflaton in the early universe and phantom DE in the late universe
even in the presence of matter. In the intermediate phase of the universe
evolution
the scalar has the standard canonical sign for kinetic energy.

\noindent 4. It is interesting to investigate the stability of the
solution (\ref{any6}). It is easier to work without matter, that
is, we put $g_0$ in (\ref{any5}) or $F$ in (\ref{any9}) to be
equal zero. One defines $d/dN\equiv \left(H^{-1}\right)d/dt$,
$X\equiv \dot\phi$, $Y\equiv f(\phi)/H$. Note that $X=Y=1$ in the
solution (\ref{any6}). By using the first FRW equation
(\ref{any1}), we find, for the solution (\ref{any6}),
\be
\label{any38}
\mu\equiv \frac{1-Y^2}{1-X^2}=\frac{H^2}{\dot H}\ .
\ee
Then by using the FRW equations (\ref{any1}), (\ref{k4}) with
(\ref{any5}) or (\ref{any9}), and the scalar field equation
\be
\label{k8}
0=\omega(\phi)\ddot \phi +
\frac{1}{2}\omega'(\phi){\dot\phi}^2 + 3H\omega(\phi)\dot\phi +
V'(\phi)\ ,
\ee
one finds
\be
\label{any39} \frac{dX}{dN}=Y-X\ ,\quad \frac{dY}{dN}=\mu X \left(1 - XY\right)\ .
\ee
Consider the perturbations from the solution $X=Y=1$: 
$X=1+\delta X$, $Y=1 + \delta Y$.  From (\ref{any39}), it follows
\be
\label{any41}
\frac{d}{dN}\left(\begin{array}{cc} \delta X \\ \delta Y
\end{array}\right) =M\left(\begin{array}{cc} \delta X \\
\delta Y\end{array}\right),\
M\equiv \left(\begin{array}{cc} -1 & 1 \\ -
\mu & -\mu \end{array}\right)\ .
\ee
If the real parts of all the
eigenvalues of the matrix $M$ are negative, the solution $X=Y=1$
is stable. The eigenvalues $\lambda_\pm$ are given by $\lambda_\pm
= \left\{-(1+\mu)\pm \sqrt{(1+\mu)^2 - 8\mu}\right\}/2$. Then the
solution (\ref{any6}) is stable if and only if $\mu>0$. From
(\ref{any38}), it is seen the positive $\mu$ means positive $\dot
H$. Hence, in the phantom phase ($\dot H>0$), the solution is
stable\footnote{
The  scalar model of deceleration/acceleration transition   has been
considered in \cite{Vikman}.
As also discussed here, the solution (\ref{any6}) with $g_0=0$
is stable in the phantom phase
but unstable in the non-phantom phase. As one of the eigenvalues $\lambda_\pm$
becomes very large when crossing $w=-1$, the instablity is high there.
In order to avoid this problem for above toy model, we may consider two
scalar model as in Appendix \ref{AA1}.
In case of one scalar model, the instability becomes infinite at the crossing $w=-1$ point,
which occurs since the coefficient of the kinetic term $\omega(\phi)$ in (\ref{k1})
vanishes at the point. In the two scalar model, one can choose the
corresponding coefficients
do not vanish anywhere. Then we may expect that such a divergence of the instability would
not occur, which can be checked in Appendix \ref{AA1}.
}.

One can check what kind of the EOS for the scalar field
appears.
For simplicity, the universe without matter is considered.
For the solution (\ref{any6}), by using the first FRW equation (\ref{any1}), we
find
\be
\label{any43}
\frac{3}{\kappa^2}H^2 = \frac{3}{\kappa^2}f(\phi)^2 = \rho\ , \quad
f(\phi)\equiv g'(\phi)\ ,
\ee
which may be solved as
$\phi=f^{-1}\left(\kappa\sqrt{\frac{\rho}{3}}\right)$.
Here $f^{-1}$ is the inverse of $f$, that is, if $y=f(x)$, then $x=f^{-1}(y)$.
By using (\ref{any2}) and (\ref{any5}) with $\rho_m=p_m=g_0=0$, it follows
$\omega(\phi)=-\frac{2}{\kappa^2}f'(\phi)=\rho + p$.
Combining the above equations,  the following
EOS may be obtained:
\be
\label{any46}
p=-\rho -
\frac{2}{\kappa^2}f'\left(f^{-1}\left(\kappa\sqrt{\frac{\rho}{3}}\right)\right)\
.
\ee
Note that $f^{-1}$ could be multi-valued function in general.
For example, in case of (\ref{any11}), EOS is
\be
\label{any47}
p=-\rho \mp \frac{(w_m + 1)\rho}{t_s}
\sqrt{t_s^2 - \frac{8}{3(w_m + 1)\kappa}\sqrt{\frac{3}{\rho}}}\ .
\ee
Similarly, EOS for other types of scalar couplings may be constructed
which shows that scalar field dynamics may be always mapped into
the (complicated) EOS.

\noindent
5. The description of unified phantom dynamics in terms of EOS
can be fitted against observations if one selects suitable sets of
data at low redshift $(z\sim 0-1)$, medium redshift $(1\ll z\ll 100)$
and extremely high redshift $(100\ll z\ll 1000)$. Specifically, observational
evidences point out that the evolutionary history of the universe
comprises two periods of accelerated expansion, namely the
inflationary epoch and the present day dark energy dominated phase
with an intermediate decelerated phase where a component of cosmic
fluid (dark/baryonic matter) has given rise to clustered large
scale structure. As we have seen, a single (effective) fluid may
indeed be responsible of both periods of accelerated expansion. At
the same time, this fluid should be subdominant during the
radiation/matter dominated epochs to give rise to baryogenesis
and structure formation. In any case, whatever the fluid is, in
order to achieve a unified model which could be matched with
observations, the cosmological energy density has to scale as
\begin{equation}
\rho(a) = {\cal{N}} a^{-3} \left ( 1 + \frac{a_I}{a} \right)^{\eta}
\left ( 1 + \frac{a}{a_{DE}} \right )^{\chi} \label{eq:rhovsa}
\end{equation}
with ${\cal{N}}$ a normalization constant, $(\eta, \chi)$ slope
parameters and $a_I \ll a_{DE}$ two scaling values of the scale
factor which lead to early inflationary epoch $(I)$ and late dark
energy epoch $(DE)$. It is convenient to rewrite Eq.(\ref{eq:rhovsa})
in terms of the redshift $z$:
\begin{equation}
\rho(z) = {\cal{N}} (1 + z)^3 \left ( 1 + \frac{1 + z}{1 + z_I}
\right )^{\eta} \left ( 1 + \frac{1 + z_{DE}}{1 + z}\right
)^{\chi} \ . \label{eq: rhovsz}
\end{equation}
To match this formalism with the above results, we have to
consider  $z = a_0/a - 1$, setting, as standard,  $a_0 = 1$ with
the subscript $0$ denoting quantities evaluated at the present
day, i.e. $z = 0$. Immediately, the Hubble rate is $H= -\frac{\dot{z}}{z+1}$.
Then, taking into account Eq.(\ref{any6}), all  results in terms
of time can be translated in terms of redshift. However, the
phenomenological parameters $\eta$ and $\chi$, assigning the slope
of $\rho$, are derived  approximating the  function $g(t)$, while
the rip time $t_s$ and the coincidence time $t_c$ are, for
each given $g(t)$-model, related respectively to $z_I$ and
$z_{DE}$. In the same way, using (\ref{any6}), the
other characteristic times as $t_a$ or $t_e$ can be translated in
terms of redshift. Keeping in mind these considerations, the
following discussion holds, in principle, for any unified phantom
cosmology, comprising dark matter, like those which we have
discussed. Essentially, $g(t)$ leads the slope of $\rho$ and the
transition between the various epochs. From Eq.(\ref{eq:rhovsa}),
it is easy to see that\,:
\begin{equation}
\left\{
\begin{array}{ll}
\rho \sim a^{-(\eta + 3)} & {\rm for} \ \ a \ll a_I \ll a_{DE} \nonumber \\
~ & ~ \nonumber \\
\rho \sim a^{-3} & {\rm for} \ \ a_I \ll a \ll a_{DE} \nonumber \\
~ & ~ \nonumber \\
\rho \sim a^{\chi - 3} & {\rm for} \ \ a_I \ll a_{DE} \ll a
\nonumber
\end{array}
\right . \ .
\end{equation}
Such an energy density scales as  dust matter in the range
$a_I\ll a \ll a_{DE}$. This means that the fluid follows matter along a
large part of the universe history, while it scales differently
only during the very beginning ($a \ll a_I$) and the present period
($a \gg a_{DE}$). Moreover, choosing $\eta = -3$, the fluid energy
density remains constant for $a \ll a_I$ thus behaving as the usual
cosmological constant $\Lambda$ during the early epoch of the
universe evolution. Finally, the slope parameter $\chi$ determines
how the fluid energy density scales with $a$ in the present epoch.
Considering an  effective EOS $w_{\rm eff} \equiv p/\rho$, we get, in
terms of redshift,
\begin{equation}
w_{\rm eff} = \frac{\eta}{3} \left( \frac{1 + z}{2 + z + z_I} \right)
  - \frac{\chi}{3} \left( \frac{1 + z_{DE}}{2 + z + z_{DE}}
\right)\,.
\label{eq: wvsz}
\end{equation}
It is worth noting that $w_{\rm eff}$ does not
depend neither on $\chi$ nor on $z_{DE}$ for high values of $z$,
that is for early epochs. On the contrary, these two parameters
play a key role in determining the behavior of the EOS over the redshift
range $(0, 100)$ which represents most of
the history of the universe (in terms of time) and the interesting
period for structure formation. The role of the different
quantities $(\eta, \chi, z_I, z_{DE})$ is better understood
considering the asymptotic limits of the EOS\,:
$\lim_{z \rightarrow \pm \infty}{w_{\rm eff}(z)} = \frac{\eta}{3}$,
%\label{eq: wlim}
where $z \rightarrow -\infty$ refers to the asymptotic future.
Setting $\eta = -3$, the fluid equation of state asymptotically
approaches that of the cosmological constant, i.e. $w_{\Lambda} =
-1$. In general, if we impose  $\eta < -1$, we get a fluid having
a negative pressure in the far past so that it is able to drive
the accelerated expansion occurring during the inflationary epoch.

It is now clear that $z_I$ controls the transition towards the
past asymptotic  value in the sense that the larger is $z$ with
respect to $z_I$, the smaller is the difference between
$w_{\rm eff}(z)$ and its asymptotic limit $\eta/3$. This consideration
suggests that $z_I$ has to take quite high values (indeed, far
greater than $10^3$) since, for $z \gg z_I$, the universe is in its
inflationary phase. The  present day value of $w_{\rm eff}$ is\,:
\begin{equation}\
w_0 = \frac{\eta}{3 (2 + z_I)} - \frac{\chi}{3} \left ( \frac{1 +
z_{DE}}{2 + z_{DE}} \right ) \simeq -\frac{\chi}{3} \left (
\frac{1 + z_{DE}}{2 + z_{DE}} \right ) \label{eq: wz}
\end{equation}
where we have used the fact that $z_I$ is very large. Being
$z_{DE} > 0$,  in order to have the present day accelerated
expansion, $w_0$ should be negative so that we get  $\chi> 0$.
Moreover, depending on the values of $\chi$ and $z_{DE}$,
$w_0$ could also be smaller than $w_{\Lambda}$ so that we may
recover phantom\,-\,like models.  The parameter $z_{DE}$ then
regulates the transition to the dark energy\,-\,like dominated
period. It is worth noting that, since $\rho$ scales with $a$ as
dust matter for a long period,  the coincidence problem is
naturally solved in  any theoretical model endowed with these
features, as the unified phantom ones presented above. Moreover,
although $w_0$ could be smaller than -1 as for a phantom field,
the equation of state asymptotically tends to a value larger than
$-1$ (provided that $-3 \le \eta \le -1$) so that  the Big Rip is
significantly delayed or does not occur at all. In order to take
into account the dust-matter-dominated era, we have to ask for an
EOS (\ref{eq: wvsz}) of the form $w_{\rm eff}(z_M) = 0$, which
gives\,:
\begin{equation}
z_M = \frac{y_{DE} \chi - (2 + y_{DE}) \eta \pm \sqrt{y_{DE}
{\cal{Z}}(\eta, \chi, z_{DE}, z_I)}}{2 \eta} \label{eq: zm}
\end{equation}
with $y_{DE} = 1 + z_{DE}$ and
\begin{equation}
{\cal{Z}}(\eta, \chi, z_{DE}, z_I) =  y_{DE} \eta^2 + 2 (2 -
y_{DE} + 2 z_I) \eta \chi + y_{DE} \chi^2 \ . \label{eq: defzm}
\end{equation}
It is easy to check that, for reasonable values of $\eta$ and
$z_I$, $z_M$ is always a complex number and hence the equation of
state  never vanishes and it is always negative. Thus, even if its
energy density scales as that of dark matter over the most of the
universe life, dark energy cannot play the same role of matter
since its equation of state is always significantly different from
null. As a result, we have to include also the dark matter in the
total energy budget. This intrinsic difference in the behavior of
dark energy and dark matter equations of state is the reason why
the former, essentially, contributes to the cosmic acceleration
while the latter gives rise to the clustered large scale
structure. This kind of analysis is particularly useful in the
case of solution (\ref{any47}) where the dominance of dark matter
or dark energy is ruled by the parameters $t_s$ and $w_m$.

Summarizing,  we have  proposed an unifying approach to the
problem of inflation, dark matter and dark energy in the same
theoretical framework. As we have seen, it is possible to
construct {\it exact} phantom-like cosmological models where
{\it all} peculiar eras of cosmic evolution are achieved. Such an
evolution can be matched with observations using the approach
outlined in \cite{salvatore} where it is considered the
dimensionless coordinate distance to Gold SNeIa sample
\cite{Riess04} and a dataset comprising 20 radio galaxies
\cite{DD04, RGdata}, the shift parameter \cite{WM04,WT04} and the
baryonic acoustic peak in the LRG correlation function
\cite{HS96}. By this approach, it is possible not only to show the
viability of the present unifying phantom model, but also to
constrain its main parameters which are, essentially, the times
(i.e. the redshift) of transition between the various epochs.  As
an independent cross check, we can also use a recently proposed
method to constrain the model parameters with the lookback time to
galaxy clusters and the age of the universe thus obtaining
consistent estimates for the model parameters \cite{FA}. In this
case we are not resorting to distance indicators but to cosmic
clocks which allow to fix with good accuracy the transition scales
(at least the one from dark matter to dark energy dominated eras).
Furthermore, the unified phantom model can be compared and
contrasted with the CMBR anisotropy and polarization spectrum,
with the data of the matter power spectrum and the growth index
\cite{WMAP,WM04,WT04}. These tests make it possible to check the
model over  different redshift ranges than the SNeIa and radio
galaxies data, offering also the possibility to tighten the ranges
for the different parameters (in particular $z_I$ and $z_{DE}$, or
  $t_{i}$ with $i=\{e,c,s,a\}$ in the above "time" description). As further
remark, this unified phantom cosmology can be related to
fundamental theories as discussed in \cite{uni} where generalized
holographic dark energy can be constructed in this scheme. In a
forthcoming paper, the detailed comparison with observations will
be developed and discussed.

\appendix

\section{Two scalar model \label{AA1}}

As seen in (\ref{any38}-\ref{any41}),
the solution (\ref{any6}) with $g_0=0$
is stable in the phantom phase
but unstable in the non-phantom phase. As one of the eigenvalues $\lambda_\pm$
of the matrix $M$ in (\ref{any41}) becomes very large when crossing $w=-1$,
the instablity is very high there.
In order to avoid this problem, we may consider two scalar model like
\bea
\label{A1}
S&=&\int d^4 x \sqrt{-g}\left\{\frac{1}{2\kappa^2}R - \frac{1}{2}\omega(\phi)\partial_\mu \phi
\partial^\mu \phi \right. \nn
&& \left.- \frac{1}{2}\eta(\chi)\partial_\mu \chi
\partial^\mu \chi - V(\phi,\chi)
\right\}\ .
\eea
Here $\eta(\chi)$ is a function of the scalar field $\chi$.
The FRW equations give
\bea
\label{A2}
&& \omega(\phi) {\dot \phi}^2 + \eta(\chi) {\dot \chi}^2 = - \frac{2}{\kappa^2}\dot H\ ,\nn
&& V(\phi,\chi)=\frac{1}{\kappa^2}\left(3H^2 + \dot H\right)\ .
\eea
Then if
\bea
\label{A3}
&& \omega(t) + \eta(t)=- \frac{2}{\kappa^2}f'(t)\ ,\nn
&& V(t,t)=\frac{1}{\kappa^2}\left(3f(t)^2 + f'(t)\right)\ ,
\eea
the explicit solution follows
\be
\label{A4}
\phi=\chi=t\ ,\quad H=f(t)\ .
\ee
One may choose that $\omega$ should be always positive and $\eta$ be
always negative, for example
\bea
\label{A5}
\omega(\phi)&=&-\frac{2}{\kappa^2}\left\{f'(\phi) - \sqrt{\alpha^2 + f'(\phi)^2}\right\}>0\ ,\nn
\eta(\chi)&=&-\frac{2}{\kappa^2}\sqrt{\alpha^2 + f'(\chi)^2}<0\ .
\eea
Here $\alpha$ is a constant. We now define a new function $\tilde f(\phi,\chi)$ by
\be
\label{A6}
\tilde f(\phi,\chi)\equiv - \frac{\kappa^2}{2}\left(\int d\phi \omega(\phi)
+ \int d\chi \eta(\chi)\right)\ ,
\ee
which gives $\tilde f(t,t)=f(t)$.
If $V(\phi,\chi)$ is given by using $\tilde f(\phi,\chi)$ as
\bea
\label{A8}
V(\phi,\chi)&=&\frac{1}{\kappa^2}\left(3{\tilde f(\phi,\chi)}^2
+ \frac{\partial \tilde f(\phi,\chi)}{\partial \phi} \right. \nn
&& \left. + \frac{\partial \tilde f(\phi,\chi)}{\partial \chi} \right)\ ,
\eea
not only the FRW equations but also the scalar field equations are also satisfied:
\bea
\label{A9}
0&=&\omega(\phi)\ddot\phi + \frac{1}{2}\omega'(\phi) {\dot \phi}^2
+ 3H\omega(\phi)\dot\phi \nn
&& + \frac{\partial \tilde V(\phi,\chi)}{\partial \phi}\ ,\nn
0&=&\eta(\chi)\ddot\chi + \frac{1}{2}\eta'(\chi) {\dot \chi}^2
+ 3H\eta(\chi)\dot\chi \nn
&& + \frac{\partial \tilde V(\phi,\chi)}{\partial \chi}\ .
\eea

In case of one scalar model, the instability becomes infinite at the crossing $w=-1$ point,
which occurs since the coefficient of the kinetic term $\omega(\phi)$ in (\ref{k1})
vanishes at the point. In the two scalar model in (\ref{A1}), the coefficients $\omega(\phi)$
and $\eta(\phi)$ do not vanish anywhere, as in (\ref{A5}). Then we may expect that such
a divergence of the instability would not occur.
We now check this explicitly in the following.

By introducing the new quantities, $X_\phi$, $X_\chi$, and $Y$ as
\be
\label{A10}
X_\phi \equiv \dot \phi\ ,\quad X_\chi \equiv \dot \chi\ ,\quad
Y\equiv \frac{\tilde f(\phi,\chi)}{H} \ ,
\ee
  the FRW  and the scalar  Eqs.(\ref{A9}) are rewritten as follows:
\bea
\label{A11}
\frac{dX_\phi}{dN}&=& - \frac{\omega'(\phi)}{2H \omega(\phi)}\left(X_\phi^2 - 1\right)
  - 3(X_\phi-Y)\ ,\nn
\frac{dX_\chi}{dN}&=& - \frac{\eta'(\chi)}{2H \eta(\chi)}\left(X_\chi^2 - 1\right)
  - 3(X_\chi-Y)\ ,\nn
\frac{dZ}{dN}&=&\frac{\kappa^2}{2H^2}\left\{X_\phi \left(X_\phi Y -1\right) \right. \nn
&& \left. + X_\chi\left(X_\chi Y -1\right)\right\}\ .
\eea
Here $d/dN\equiv H^{-1}d/dt$. For the solution (\ref{A4}),
$X_\phi=X_\chi=Y=1$.
The perturbations are considered as
\be
\label{A12}
X_\phi=1+\delta X_\phi\ ,\quad X_\chi=1 + \delta X_\chi\ ,\quad Y=1 + \delta Y\ .
\ee
Then
\bea
\label{A13}
&& \frac{d}{dN}\left(\begin{array}{c}
\delta X_\phi \\
\delta X_\chi \\
\delta Y
\end{array}\right)
= M \left(\begin{array}{c}
\delta X_\phi \\
\delta X_\chi \\
\delta Y
\end{array}\right)
\ ,\nn
&& M\equiv \left(\begin{array}{ccc}
- \frac{\omega'(\phi)}{H\omega(\phi)} - 3 & 0 & 3 \\
0 & - \frac{\eta'(\chi)}{H\eta(\chi)} - 3 & 3 \\
\frac{\kappa^2}{2H^2} & \frac{\kappa^2}{2H^2} & \frac{\kappa^2}{H^2}
\end{array}\right)\ .
\eea
The eigenvalues of the matrix $M$ are given by solving the following eigenvalue
equation
\bea
\label{A14}
0&=& \left(\lambda + \frac{\omega'(\phi)}{H\omega(\phi)} + 3\right)
\left(\lambda + \frac{\eta'(\chi)}{H\eta(\chi)} + 3\right) \nn
&& \times \left(\lambda - \frac{\kappa^2}{H^2}\right) \nn
&& + \frac{3\kappa^2}{2H^2}\left(\lambda + \frac{\omega'(\phi)}{H\omega(\phi)} + 3\right) \nn
&& + \frac{3\kappa^2}{2H^2}\left(\lambda + \frac{\eta'(\chi)}{H\eta(\chi)} + 3\right)\ .
\eea
The eigenvalues are clearly finite. Then even if there is an instability,
it could be finite. More complicated models along this line may be
presented as well.

\end{document}